# Estimation of Young's Modulus of Graphene by Raman Spectroscopy

*Jae-Ung Lee, Duhee Yoon, and Hyeonsik Cheong\**

Department of Physics, Sogang University, Seoul 121-742, Korea.

**AUTHOR EMAIL ADDRESS**: jaeunglee@sogang.ac.kr; dhyoon@sogang.ac.kr; and hcheong@sogang.ac.kr

**CORRESPONDING AUTHOR FOOTNOTE:** E-mail: hcheong@sogang.ac.kr; TEL: 82-2-705-8434; Fax: 82-2-717-8434.

**ABSTRACT:** The Young's modulus of graphene is estimated by measuring the strain applied by a pressure difference across graphene membranes using Raman spectroscopy. The strain induced on pressurized graphene balloons can be estimated directly from the peak shift of the Raman *G* band. By comparing the measured strain with numerical simulation, we obtained the Young's modulus of graphene. The estimated Young's modulus values of single- and bi-layer graphene are 2.4±0.4 TPa and 2.0±0.5 TPa, respectively.

The mechanical properties of graphene are interesting research subjects because its Young's modulus and strength are known to be extremely high. Owing to these properties, graphene is a promising candidate for applications in nanomechanical systems.[1–3] Determination of such mechanical properties of graphene in nanometer scale is an important issue not only for applications in nanomechanics but also in studying fundamental physical properties. Values of ~1 TPa for the Young's modulus have been reported in mechanically exfoliated graphene samples by the nanoindentation technique using an atomic force microscopy and by the constant-N blister test.[4,5] In this work, we take a different approach and estimate the Young's modulus of graphene by measuring directly the strain applied to a graphene membrane under a differential pressure using Raman spectroscopy. Since graphene is impermeable to any gas,[6] it can sustain a differential pressure of several bars.[5–7] By applying such a differential pressure on a graphene membrane held over a circular aperture, a graphene balloon is formed with a biaxial strain applied at the center of the circular membrane. Raman spectroscopy is a very powerful tool for investigating the intrinsic properties of graphene.[8–17] Especially, the Raman spectrum of graphene is very sensitive to mechanical deformations. Recent experiments demonstrated that both the Raman $G$ and $2D$ bands red-shift and split into two peaks each under uniaxial strain.[18–21] Therefore, strain on a graphene membrane can be measured by Raman spectroscopy very accurately. By measuring the strain induced on pressurized graphene balloons by Raman spectroscopy and comparing the strain with a numerical calculation based on the finite element method, we could deduce the Young's modulus of graphene layers. This method can provide valuable information on the intrinsic properties of graphene and can be applied to other 2-dimensional materials for investigating their mechanical properties.

Graphene samples were prepared on pre-patterned silicon substrates covered with a 300-nm thick $SiO_2$ layer. The substrates were patterned with round holes by photolithography and dry etching. The depth of a hole is ~5 μm, and the diameters are 2.0, 3.1, 4.2, 5.3, 6.4, and 7.3 μm. The samples were prepared directly on cleaned substrates by mechanical exfoliation from natural macrocrystalline graphite flakes (NGS Naturgraphit GmbH, Germany). The samples were placed into a vacuum chamber to make pressure difference across the graphene membrane.

The Raman measurements were performed with the 514.5-nm (2.41 eV) line of an Ar ion laser. The laser beam was focused onto the graphene sample by a 40× microscope objective lens (N.A. 0.65) through a quartz window. The scattered light was collected and collimated by the same objective, dispersed with a Jobin-Yvon Triax 550 spectrometer (1800 grooves/mm), and detected with a liquid-nitrogen-cooled charge-coupled-device detector. The spectral resolution was about 0.7 cm$^{-1}$. The focused beam size was measured with the razor-edge method and the spatial resolution was ~1 μm. The laser intensity was kept below 0.5 mW to avoid local heating induced by the laser. For the Raman scanning image, a spectrum was measured at each position of the sample, raster-scanned in a rectangle with 30×30 points in 0.3-μm steps.

Figure 1(a) shows the schematic diagram of the experimental setup. In the vacuum chamber, a pressure difference across the graphene membrane is applied by pumping out the vacuum chamber. This pressure difference makes the graphene membrane bulge upward like a balloon. A scanning electron microscope (SEM) image of a sample measured in vacuum is shown in Figure 1(b). Because samples tend to be damaged by the electron beam, the samples used for the SEM measurements were from a different batch than those for Raman measurements. This image clearly shows that the graphene membrane bulges upward. The deflection of the graphene on a 6.6 μm diameter hole estimated from the SEM image is about ~200 nm, which is consistent with a numerical simulation using the finite element method.

Raman spectroscopic measurements were performed on the single-layer graphene sheet shown in Fig. 2(a). The single Lorentzian shape of the 2$D$ band in Fig.2(c) in the supported region indicated that the sample is clearly a single layer.[22,23] But in the suspended region, some asymmetry of the 2$D$ band is observed. It is consistent with the previously reported data for the suspended single layer graphene sample.[24] At atmospheric pressure, the $G$ peak position in the supported region outside the hole is substantially blue-shifted with respect to the suspended region (Fig. S1), and the integrated intensity ratio of the Raman 2$D$ to $G$ band is about 4. These and the narrow linewidth of the $G$ peak indicate that the supported graphene is highly doped.[10,11] It is well known that graphene on SiO$_2$/Si substrates can be strongly doped.[12] At the center of the hole, the peak position of the $G$ band is about 1581 cm$^{-1}$ and the intensity of Raman $G$ band is very small relative to that of the 2$D$ band, with the intensity ratio of Raman 2D to G band ~11, which indicates that the suspended part of the graphene sample is minimally doped.[24,25] Furthermore, the peak position indicates that the initial residual tension induced by Van der Waals force between graphene and edges of the hole is very small.[6] This implies that our sample is suitable for investigating the intrinsic properties of undoped graphene.

When the chamber is evacuated, there are no distinctive changes in the supported region. On the other hand, the Raman spectrum from the center of the suspended graphene is red-shifted in vacuum [Fig. 2(c)]. The G peak position at the center of the hole is shifted to ~1568 cm$^{-1}$. Figure 2(b) is an image of the G peak position of the graphene sample over a 6.4-µm hole in vacuum. In the supported region the frequency of the G peak is near 1591 cm$^{-1}$. The G peak position gradually red-shifts as the probing position moves to the center of the hole with the minimum value of ~1568 cm$^{-1}$. This red-shift is caused by the tension due to bulging of the graphene membrane. Once the evacuation is complete, the Raman spectrum did not change after more than 2 hours in vacuum, and when the pressure is returned to 1 bar by letting air into the vacuum chamber, the G peak position at the center went back to 1581 cm$^{-1}$ (Fig. S2). The G peak position was reproducible after many cycles of evacuation and vacuum release. This is evidence that there is no appreciable leakage of the gas confined in the hole. This result is consistent with previous reports.[5,6]

By measuring the shifts of the Raman G and 2D bands, we can estimate the magnitude of the strain on a graphene membrane.[18–21] To estimate the strain from the Raman spectrum, we use the reported value of the Grüneisen parameter ($\gamma$) and the shear deformation potential ($\beta$). The used values are $\gamma = (1/\omega_0)(\partial \omega / \partial \varepsilon_h) = 2.2 \pm 0.1$ and $\beta = (1/\omega_0)(\partial \omega / \partial \varepsilon_s) = 0.93 \pm 0.04$ [20] where $\omega_0$ is the phonon frequency of unstrained graphene, and $\varepsilon_h$ and $\varepsilon_s$ are the hydrostatic and shear components of the strain, respectively. Using these values, one obtains the G phonon shift due to biaxial strain $\varepsilon_b$, $\Delta \omega_b = -2\omega_0 \gamma \varepsilon_b$, to be 70 cm$^{-1}$ per biaxial strain of 1%.[16] The G peak shift of ~13 cm$^{-1}$ for the 6.4-µm hole corresponds to a biaxial strain of ~0.19% at the center of the graphene membrane. We measured the changes of the Raman spectrum at the center of the hole with various diameters. Figure 3 shows that both the Raman *G* and *2D* bands are red-shifted and the shifts are dependent on the diameter of the hole. The *G* peak shifts are 5.0±3.6, 8.0±2.8, 10.0±1.3 and 13.0±1.0 cm$^{-1}$ for the diameters of 3.1, 4.2, 5.3 and 6.4 μm, respectively. These values in turn correspond to biaxial strain values of 0.07±0.05, 0.11±0.04, 0.14±0.03 and 0.19±0.02%, respectively. The biaxial strain on a graphene balloon increases as a function of the size of the hole. Some asymmetry is observed in the line shape of the *G* band from the smallest 3.1-μm hole. Since the laser spot size, ~1 μm, is a significant fraction of the total hole size, the strain variation within the laser spot is large relative to the maximum strain at the center. This strain variation causes asymmetric line shapes for smaller holes, which contributes to larger error bars.

We compared the obtained strain values with a numerical simulation based on the finite element method. The graphene membrane was modeled by a clamped circular membrane with a linear elasticity. It is well known that graphene has a non-linear elasticity[26–28] with a form of $\sigma = E\varepsilon + D\varepsilon^2$, where *E* is the Young's modulus, *D* the third-order elastic modulus, $\sigma$ the stress, and $\varepsilon$ the uniaxial strain.[4] Since typical values of *E* and *D* are 1.0 TPa and –2.0 TPa,[4] respectively, and the maximum strain applied in our experiments was ~0.2%, the first term is much larger than the second term, and so the elastic behavior of graphene may be assumed to be linear in our experimental conditions. The confined air inside the hole under the graphene membrane is initially at atmospheric pressure. When the vacuum chamber is evacuated to a pressure below 10$^{-5}$ torr, the graphene membrane bulges upward and as a result, the volume of the confined gas increases. Because of this effect the pressure difference across the membrane drops slightly. If one uses the deflection of ~200 nm estimated from the SEM image [Fig 1(b)], the pressure difference is ~0.96 atm. This value was used as the initial estimate of the pressure difference in the iterative procedure to find the Young's modulus. With the iterative fitting procedure, the deflection converges to ~160 nm and the pressure difference to 0.97 atm. We assumed that the Poisson's ratio of graphene is the same as that of graphite, 0.16.[29] For the thickness of the graphene membrane, we used the interlayer spacing of graphite, 0.335 nm,[30] and ignored the small thickness change by the deflection.

Numerical simulations were performed using a commercial finite element program, ABAQUS. The diameter of the circular membrane is taken to be the same as the diameter of the hole. A uniform pressure was applied perpendicular to the membrane. As a boundary condition, the edges of the membrane were clamped to the circular edge of the hole. This is reasonable since Koenig *et al.* recently reported that the graphene membranes firmly adhere to the substrate up to a pressure difference of >2.5 MPa (25 atm).[5]

If we use the previously reported Young's modulus value[4,5] of 1 TPa in our simulation, the deflection is ~220 nm and the strain at the center of the circular membrane is ~ 0.32% for the 6.4 μm hole. This is much larger than the measured value of 0.19%. If we instead use the Young's modulus as a fitting parameter to reproduce the measured strain at the center, we obtain a Young's modulus value of 2.4 TPa. To further confirm the consistency of our analysis, we repeated for the different hole diameters. We obtained Young's modulus values of 2.7±1.2, 2.7±0.8, 2.5±0.4 and 2.4±0.3 TPa for diameters of 3.1, 4.2, 5.3 and 6.4 μm, respectively. Our estimation of the Young's modulus of single-layer graphene is 2.4±0.4 TPa.

Figure 4 shows the calculated strain distribution on the circular membrane using the Young's modulus value of 2.4 TPa. The strain in the transverse direction does not vary much whereas the radial strain varies a lot. At the center, a biaxial strain of 0.19% is reproduced. Near the edge of the membrane, the transverse and radial strains are significantly different, resulting in a shear component. We attempted to detect this shear component using polarized Raman spectroscopy.

Figure 5(a) shows the change of the *G* peak position along the diameter of the graphene membrane, measured in two orthogonal polarizations. The incident laser is polarized in the vertical direction and the laser spot position is moved horizontally. The analyzer for the scattered signal was set either vertically (0°) or horizontally (90°). The *G* peak position is independent of the polarization near the center whereas a small shift is seen near the edges. Figure 5(b) compares the Raman spectra for the two polarizations at different positions. At the center or at 1 μm from the center, the two spectra are almost identical. If we use the known splitting of the *G* peak due to uniaxial strain, the splitting would be 0.2 cm$^{-1}$ corresponding to a uniaxial strain of 0.01% at 1 μm from the center, which is much smaller than the width of the *G* peak. At the position of 2 μm away from the center, a small relative shift of the *G* peak for the two polarizations is observed. The uniaxial component of the strain at this position is 0.05% from our simulation. This would give a *G* peak splitting of 1.0 cm$^{-1}$, which is consistent with what is seen in Fig. 5(b). At 3 μm away from the center, the laser spot (~1 μm) covers both the suspended and the supported regions, resulting in larger linewidths and asymmetric line shapes.

We also performed the same experiment on a bilayer graphene sample. The bilayer graphene sample fully covered a 7.3-μm diameter hole. When the vacuum chamber is evacuated, the *G* peak position is shifted from 1581 cm$^{-1}$ to 1570 cm$^{-1}$, which corresponds to a biaxial strain of 0.14% (Fig. 6). Using the same procedure as for the single layer sample, we found that the Young's modulus value for bilayer graphene is 2.0±0.5. The Young's modulus values for single- and bi-layer graphene, 2.4±0.4 TPa and 2.0±0.5 TPa, respectively, can be compared with that of graphite 1.02 TPa.[29] (The data for a 4-layer graphene sample is also included in Supporting Information.) This seems reasonable if one considers the fact that the Grüneisen parameters for graphene and graphite are $\gamma = 2.2 \pm 0.1$ and 1.06, respectively.[20,31] Since the Grüneisen parameter is closely related to elastic properties, a correlation between the Grüneisen parameter and the Young's modulus is expected.

Our estimations for the Young's modulus are significantly larger than the reported values in the literature, ~1 TPa.[4,5] A probable explanation is that the Young's modulus may not be constant in different strain ranges. For example, Lee *et al.* fitted their data mostly in the range of 0 to 5% of strain with a model assuming a linear behavior to obtain the Young's modulus value of 1.0 TPa for single layer graphene. Koenig *et al.* on the other hand, fitted their data in the range of 0.25 to 0.5 MPa (2.5 to 5 atm) with a linear model to obtain the Young's modulus for 1-5 layer graphene. In both cases, if graphene has a significant softening at higher strain ranges, the estimated Young's modulus could be smaller than the value estimated in the small strain range. In our work, the maximum strain was only 0.19%, at least an order of magnitude smaller than the maximum strain in previous work. The behavior at even lower strain was inspected by repeating the measurements as a function of the pressure in the vacuum chamber between 0 and 1 atm. (See Figures S4 and S5 in Supporting Information.) Although accurate quantitative analysis is not possible due to the large uncertainty caused by small Raman peak shifts, the estimated Young's modulus value is consistently larger than the previous reports.

In conclusion, we found that the Young's modulus values of single- and bi-layer graphene are 2.4±0.4 TPa and 2.0±0.5 TPa, respectively. Comparison with previous estimates suggests that the linear Young's modulus value may depend on the strain range and is larger in small strain ranges.

**ACKNOWLEDGMENT**: We thank H. Kim and S. W. Lee for the SEM measurements. This work was supported by the National Research Foundation (NRF) grants funded by the Korean government (MEST) (No. 2011-0013461 and No. 2011-0017605). This work was also supported by a grant (No. 2011-0031630) from the center for Advanced Soft Electronics under the Global Frontier Research Program of MEST.

**Supporting Information Available:** Raman *G* peak position as a function of the position along the center line of the hole, the repeatability data, Raman spectra using different excitation energies, Raman spectra measured at different pressures (0~1 bar), Young's modulus estimated at different pressures, Raman spectra for a 4-layer sample, and Young's modulus as a function of the number of layers. This material is available free of charge via the internet at http://pubs.acs.org.

**FIGURE CAPTIONS**

**Figure 1.** (a) Schematic diagram of the experimental setup. (b) Scanning electron microscope (SEM) image of a graphene membrane in vacuum, taken at an oblique angle. Two bubbles with different diameters (3.6 and 6.6 μm), formed by the pressure difference between inside and outside of the hole, are shown.

**Figure 2.** (a) Optical microscope image of a graphene sample on a pre-patterned $SiO_2$/Si substrate. The bright region corresponds to a thick part of the sample, and the faintly darker area next to it corresponds to the single layer region. (b) Raman map of the *G* peak position of a single-layer graphene balloon on the 6.4 μm hole. (c) Raman spectra measured at the center of the 6.4 μm hole and at a supported region of a single-layer sample.

**Figure 3.** Comparison of the Raman (a) *G* band and (b) *2D* band spectra, measured from a single-layer sample in vacuum and in ambient pressure at the center of the holes with 3.1, 4.2, 5.3 and 6.4 μm diameters.

**Figure 4.** Simulated strain distribution over the single-layer graphene membrane obtained by numerical calculations. (Inset) Comparison of the radial and transverse components of the strain distribution

**Figure 5**. (a) Shift of the *G* peak position due to the pressure difference as a function of the position along the center of the hole, measured in two orthogonal polarizations. (b) Raman *G* band spectra for two orthogonal polarizations, measured at different distances from the center.

**Figure 6.** Comparison of the Raman spectra, measured from a bilayer sample in vacuum and in ambient pressure at the center of a 7.3-μm hole and at a supported region.

26. Liu, F.; Ming, P. B.; Li, J. *Phys. Rev. B* **2007**, 76, 064120.

27. Wei, X.; Fragneaud, B.; Marianetti, C. A.; Kysar, J. W. *Phys. Rev. B* **2009**, 80, 205407.

28. Jiang, J.; Wang, J.; Li, B. *Phys. Rev. B* **2010**, 81, 073405.

29. Blakslee, O. L.; Proctor, D. G.; Seldin, E. J.; Spence, G. B.; Weng, T. *J. Appl. Phys.* **1970**, 41, 3373–3382.

30. Aljishi, R.; Dresselhaus, G. *Phys. Rev. B* **1982**, 26, 4514 –4522.

31. Hanfland, M.; Beister, H.; Syassen, K. Phys. Rev. B 1989, 39, 12598–12603.12

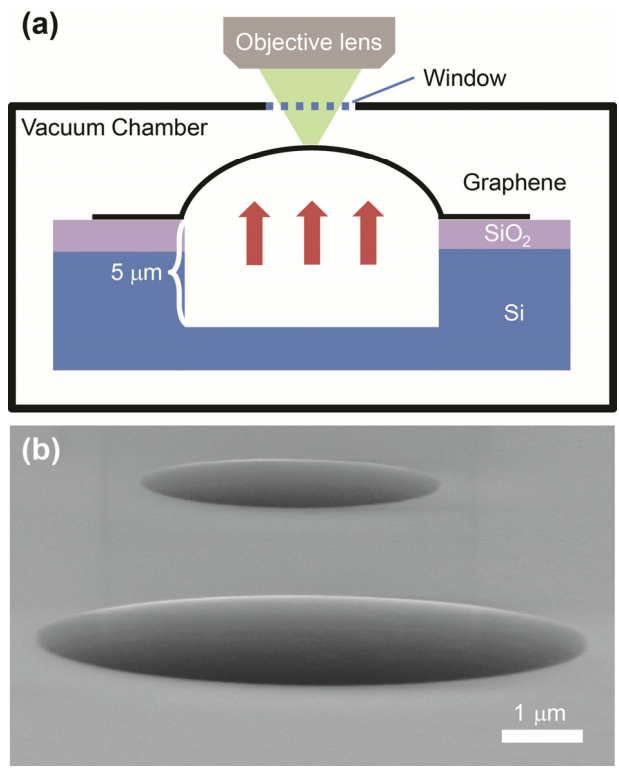

**FIGURE 1**

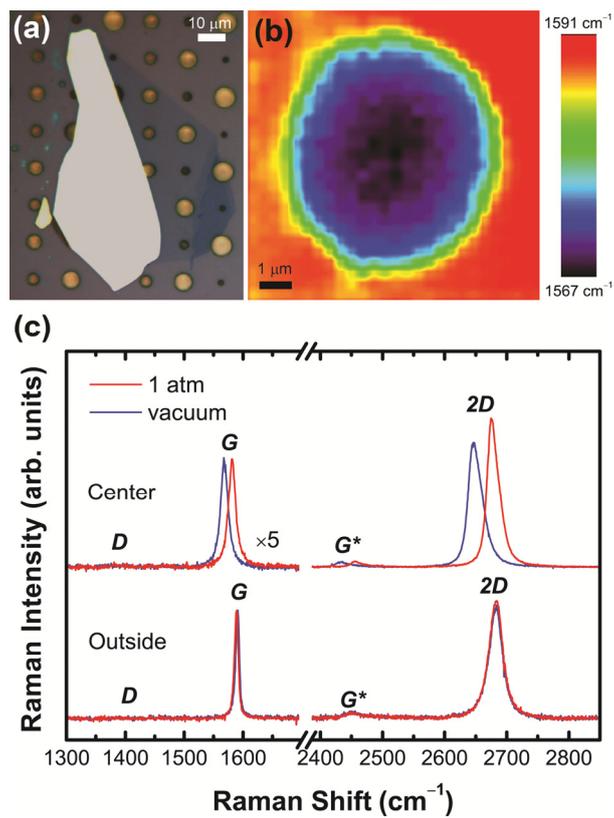

**FIGURE 2**

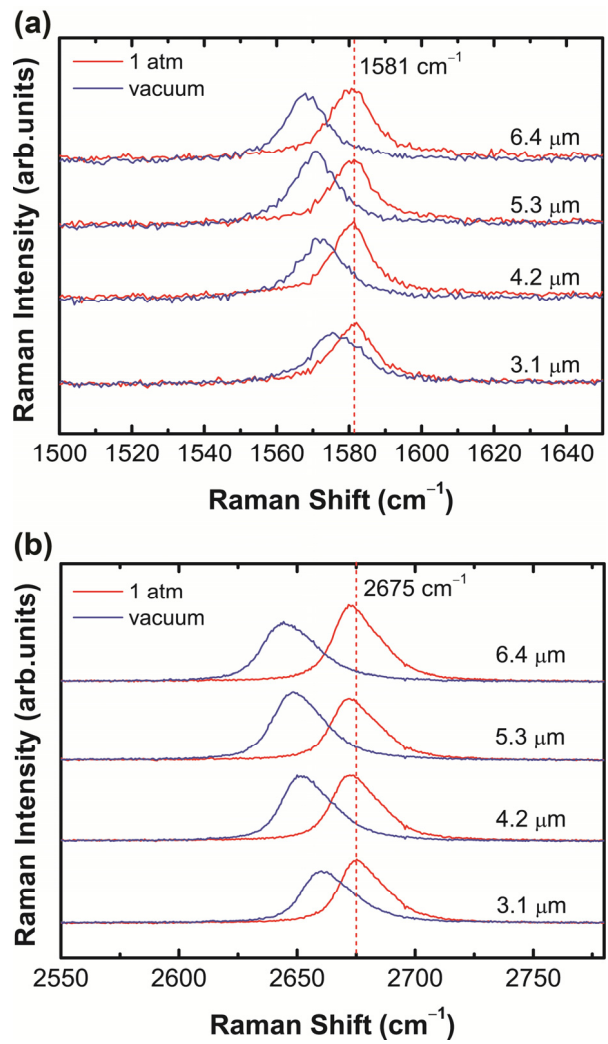

**FIGURE 3**

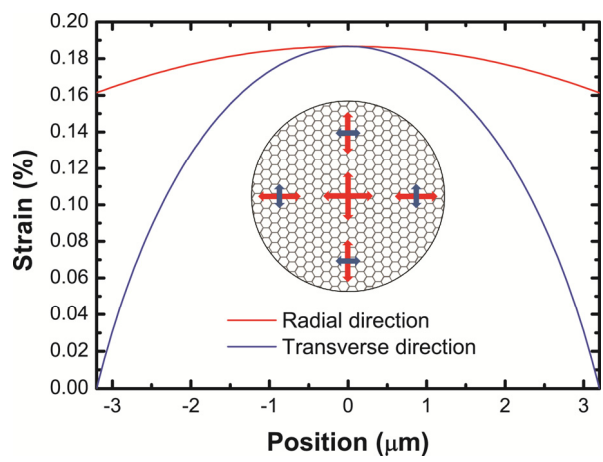

**FIGURE 4**

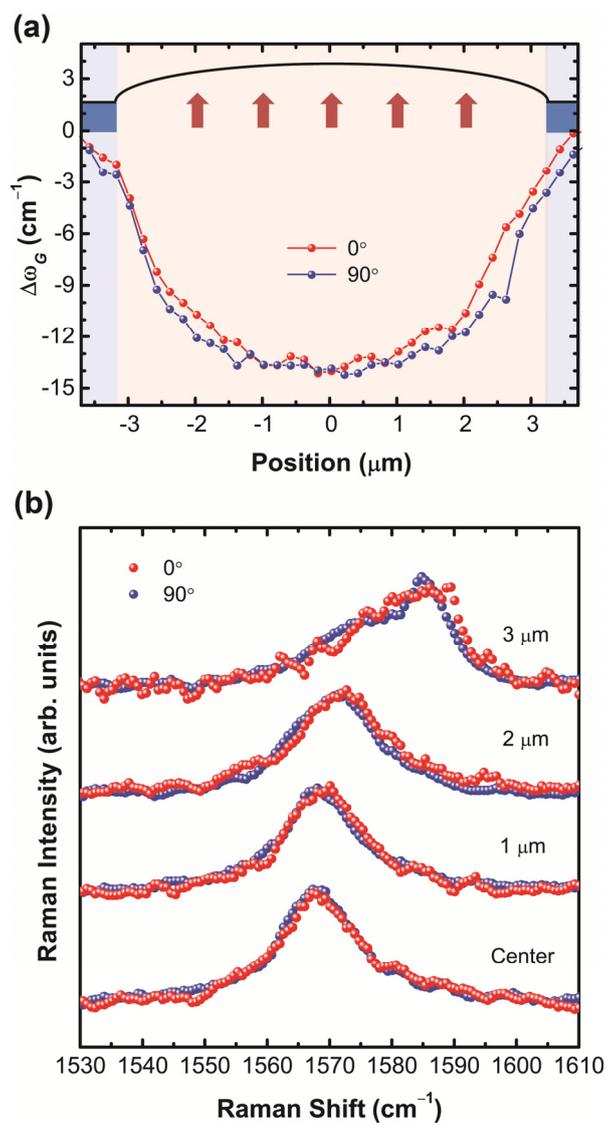

**FIGURE 5**

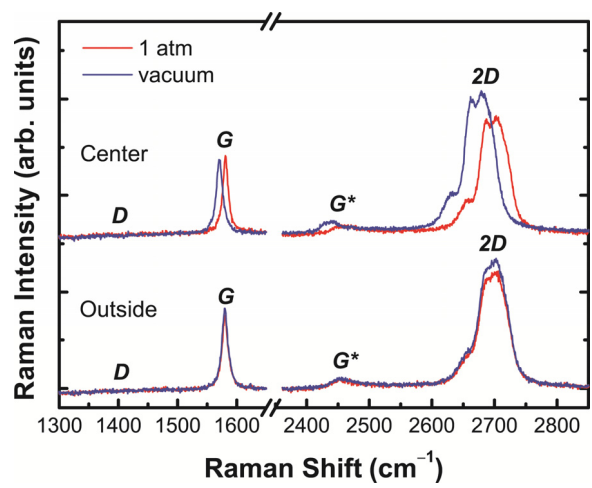

**FIGURE 6**

**SYNOPSIS TOC**

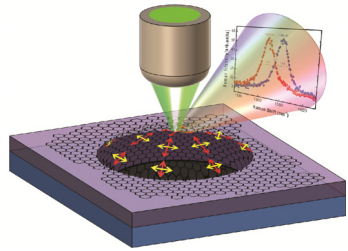

# Estimation of Young's Modulus of Graphene by Raman Spectroscopy

*Jae-Ung Lee, Duhee Yoon, and Hyeonsik Cheong*[*]

Department of Physics, Sogang University, Seoul 121-742, Korea

**AUTHOR EMAIL ADDRESS:** jaeunglee@sogang.ac.kr, dhyoon@sogang.ac.kr; hcheong@sogang.ac.kr

**CORRESPONDING AUTHOR FOOTNOTE:** E-mail: hcheong@sogang.ac.kr; TEL: 82-2-705-8434; Fax: 82-2-717-8434

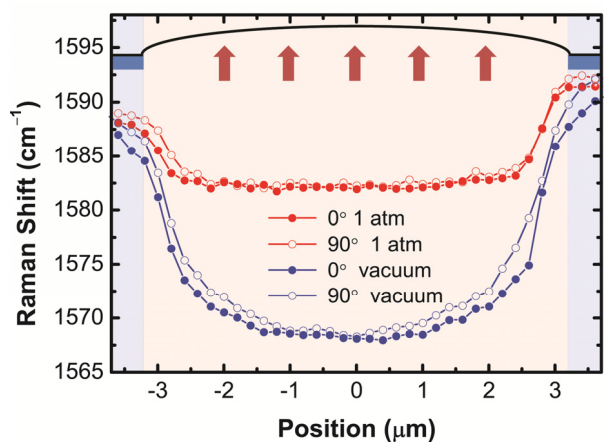

**Figure S1.** Position of the Raman *G* band as a function of the distance from the center of the hole of a single-layer sample, measured in two orthogonal polarizations in atmospheric pressure (red) and vacuum (blue).

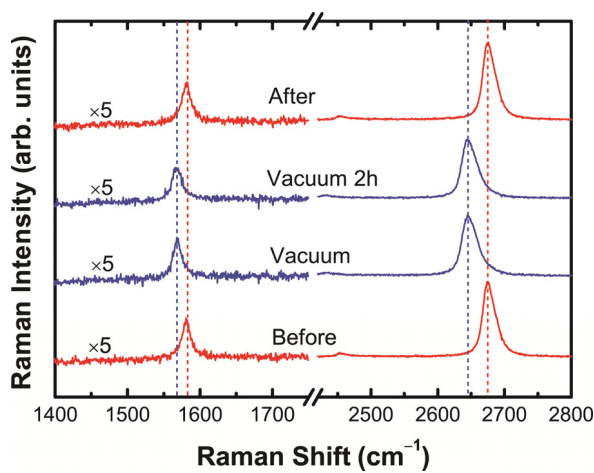

**Figure S2.** Evolution of the Raman spectrum taken at the center of the hole of a single-layer sample: before the vacuum chamber is evacuated (Before), right after the chamber is evacuated (Vacuum), after the sample was held in vacuum for 2 hours (Vacuum 2h), and after the chamber pressure was raised to atmospheric pressure (After). This illustrates that the pressure difference is maintained for at least 2 hours in vacuum, and there is minimal escape of the gas from the graphene balloon during the cycle.

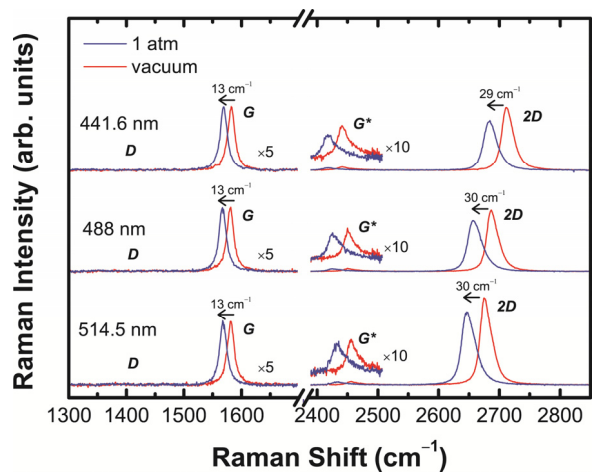

**Figure S3.** Raman spectra measured at the center of the 6.4 μm hole at 1 atm. (unstrained) and in vacuum (strained). Three different laser lines were used (441.6-nm line of a He-Cd laser, 488- and 514.5-nm lines of an Ar ion laser). The peak shift is almost the same for all laser lines.

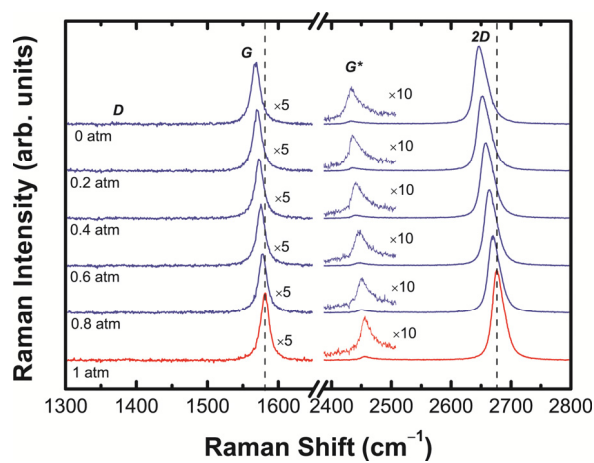

**Figure S4.** Raman spectra measured at the center of the 6.4 μm hole at different pressures.

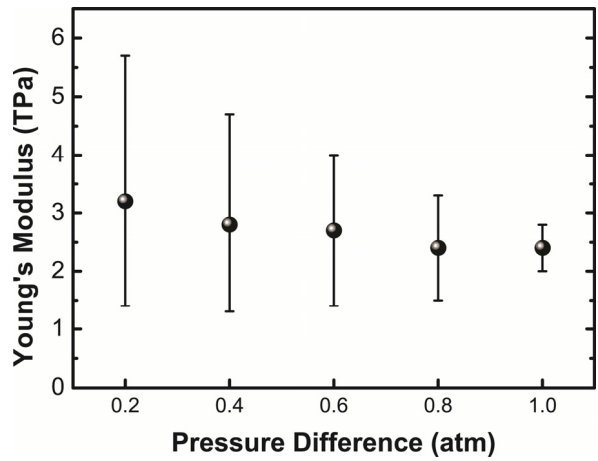

**Figure S5.** Young's modulus values estimated from Raman measurements at different pressure differences (Fig. S4).

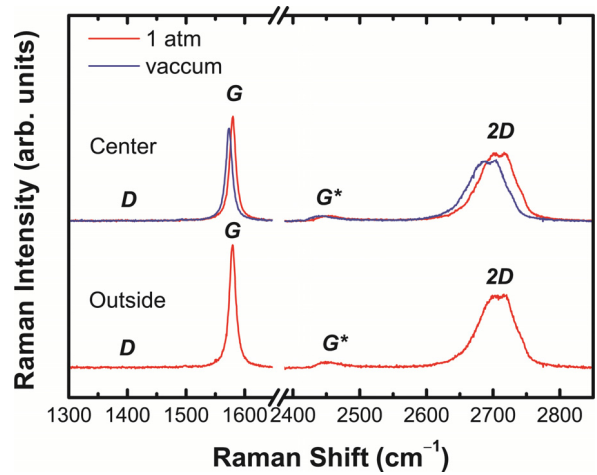

**Figure S6.** Comparison of the Raman spectra, measured from a 4-layer sample in vacuum and in ambient pressure at the center of a 7.3-µm hole and at a supported region.

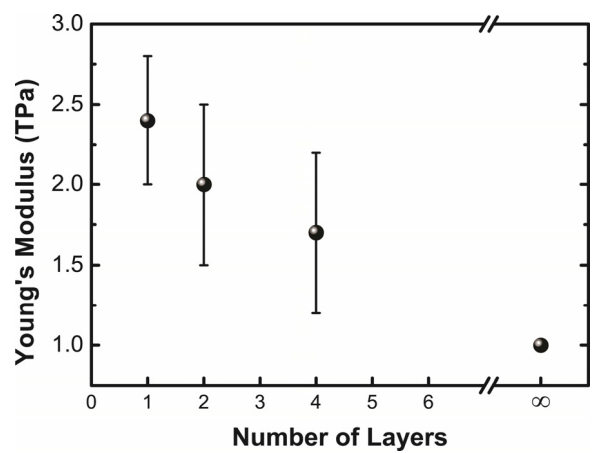

**Figure S7.** Estimated Young's modulus as a function of the number of layers. Young's modulus decreases as the number of layers increases.